\documentstyle[12pt]{article}
\begin{document}
\def\bfx{\mbox{\boldmath$x$}}
\par\noindent
To appear in Acta Physica Polonica B\\
hep-ph/9606263 \\

\begin{center}
{\bf DCC: ATTRACTIVE IDEA SEEKS SERIOUS CONFIRMATION}\\ \vspace{0.8cm}
J.-P. BLAIZOT $^a$ and A. KRZYWICKI $^b$\\ 
\vspace{0.7cm}    
$^a$ Service de Physique 
Th\'eorique\footnote{Laboratoire de la Direction des
Sciences de la Mati\`ere du Commissariat \`a l'Energie
Atomique}, CE Saclay, 91191 Gif s/Y, France\\
$^b$ Laboratoire de Physique Th\'eorique et Hautes Energies, 
Universit\'e de Paris-Sud, 91405 Orsay, 
France\footnote{Laboratoire associ\'e au CNRS.}
\vspace{1cm}\par
{\em Dedicated to Andrzej Bialas in honour of his 60th birthday}
\end{center}
\vspace{1cm}
The theoretical ideas relevant for the physics of the
disorientend chiral condensate (DCC) are reviewed.

\vspace{0.7cm}\par\noindent
PCAC numbers: 25.75.-q, 13.85.Hd\\

\vspace{4cm}\par\noindent
June 1996\\
LPTHE Orsay 96/37
\newpage
\section{Introduction} 
As is well known, quantum chromodynamics 
(QCD), the theory of 
strong interactions, has an approximate 
global SU(2)$ \times $ SU(2) 
invariance. This invariance is spontaneously 
broken and the relevant 
part of the order parameter is a 
vector $\phi = (\sigma, \vec{\pi})$ 
transforming under the O(4) subgroup 
of SU(2)$\times $SU(2). In the 
physical vacuum $\phi$ points in the 
$\sigma$ direction. One calls {\em 
disoriented chiral condensate} (DCC) a medium 
where $\phi$ is {\em coherently misaligned}. 
Experimental observation of 
a signal of transient DCC formation 
would be a striking probe of 
the chiral phase transition. 
\par
DCC is now a topical subject.
This research has a prehistory
[1-6]: Some results
have already been found long ago, but forgotten 
for various reasons, mostly because they were apparently 
lacking theoretical underpinning. They have been 
rediscovered independently in the new 
context, with a better motivation.
Actualy the true history of DCC 
research begins in the early 90's, 
when several points have been simultaneously realized:
\par
$\bullet$  In spite of the fact that 
confinement and hadronization are
quantum  phenomena, the production of 
multiparticle {\em hadron}  states can
be described in terms of  an  {\em effective} 
theory for which a classical
approximation is meaningful. In practice, 
one is led to consider classical
radiation  of soft pions, described by the
$\sigma$-model, in very high  multiplicity 
events, where this radiation is
intense (the most  promising applications are in high-energy 
heavy-ion collisions) \cite{ans,bj,bk1}. 
\par
$\bullet$ It is rather natural to expect that in some 
high-energy collisions there could appear a space region, shielded 
for some time from the outer physical vacuum, where a DCC can 
develop \cite{bj}. In particular, this is what one finds in 
the solution of the $\sigma$-model corresponding
to Heisenberg's idealized boundary conditions \cite{bk1} 
\footnote{This solution belongs to a general class of
solutions proposed earlier in \protect\cite{ans}.}.
\par
$\bullet$ It is a common feature of 
the non-trivial solutions of
non-linear field equations with internal 
symmetries, that they break the symmetry.
In the present case one expects the field 
configuration to break the O(4)
symmetry.  Since there is no a priori reason 
to priviledge any direction  in the
internal space, all field orientations 
are expected to be equiprobable. Assuming
that the isospin orientation of the pion 
field is constant throughout a space
domain - remember, that we are interested in 
long wavelength modes - and points
in some  random direction, one predicts
that the ratio
\begin{equation}
f = {{N_{\pi^0}} \over {N_{\pi^0} + N_{\pi^-} + N_{\pi^+}}} 
\label{f}
\end{equation}
should be distributed according to the simple law
 \footnote{Apparently this result appeared for the first 
time in \cite{and}, where the distribution of the neutral 
pion fraction has been calculated
for a coherent multipion state of total isospin zero. 
A simple quasi-classical argument, given first in 
\protect\cite{bk1}, enables one
to find (\protect\ref{law}) immediately: The intensity of pion 
radiation is quadratic in the pion field. 
Hence, one seeks the probability
that a unit vector, randomly oriented in isospace, has its 
3rd component equal to $\sqrt{f}$. The calculation is elementary.}
\begin{equation}
dP(f) = {{df} \over {2\sqrt{f}}}
\label{law}
\end{equation}
Hence, DCC formation has a very distinctive signature,
which actually reflects correlations coded in the
classical, i.e. coherent field. 
The deviation from the narrow Gaussian centered at
 $f = {1 \over 3}$ predicted by statistical arguments is striking.
\par
$\bullet$ Long wavelength
modes of the pion field can be
dramatically amplified during the out-of-equilibrium cooling 
of the quark-gluon plasma \cite{rw1,rw2}. 
\par
We shall develop these points in the
following sections. Our aim is to explain some of the main 
ideas, not to give a comprehensive account of 
all papers  on this rapidly
evolving subject.  
Reader's attention is called to some earlier reviews.
Those of Bjorken and collaborators [12-14]
are particularly helpful in gaining physical intuition.
They also contain a discussion of relevant experimental 
matters. A comprehensive
presentation of the subject can be found in an 
excellent review by Rajagopal \cite{raj}
 \par
\section{Choice of theoretical framework} 
For a DCC to
be produced, there must
be a stage, during the collision process,
where chiral symmetry is restored.
It is tempting to identify this stage
with the formation of a hot quark-gluon
plasma. The latter can be directly described 
in QCD language, and insight into the 
corresponding physics can be obtained using
the perturbative aproach.
The later stages of the collision involve 
eventually soft pions radiation.
This can be desribed by an effective theory 
in which the Lagrangian has the
form of a series involving an increasing 
 number of derivatives. The first
term, the one with two derivatives,
is uniquely determined. It corresponds
to the so-called non-linear
$\sigma$-model and gives account of the
 physics of the softest modes. An educated guess
is needed to figure out what
exactly happens at the intermediate stage between
these two extremes, i.e.
to describe the cooling  of the  plasma 
leading to its decay into pions.
\par
One commonly uses for the description of 
this intermediate stage the linear
$\sigma$-model. There are several reasons 
for this choice, which we
briefly discuss. First of all, the 
chiral symmetry is spontaneously 
broken when
one crosses the phase transition point. 
There exist suggestive arguments
\cite{pw,rw1} to the effect that QCD with two flavors of
massless quarks belongs to the same
{\em static} universality class as an
O(4) Heisenberg ferromagnet. Hence,
in the static regime, the long wavelength modes
can be described by a Ginzburg-Landau Lagrangian, 
which in this case is
identical to that of the linear $\sigma$-model. 
Integrating out the $\sigma$ field in the latter
one gets the non-linear $\sigma$-model Lagrangian 
as the leading term, plus higher derivative
corrections. Thus, the linear $\sigma$-model offers a correct 
description of very long wavelength pion modes. 
Furthermore, it can be, at least formally, extended 
to describe the disordered state, where all four 
components of $\phi$ are fluctuating
independently. 
\par
For these reasons, the linear $\sigma$-model appears as a
 natural first choice in the modelling of DCC formation and 
dynamics \footnote{One should mention that
interesting studies of alternative models have
also been presented. We shall
not enter here into the discussion of these
models \protect\cite{bed,bccmpg}. The $\sigma$-model 
will suffice to illustrate all the relevant 
points.}. The following caveat should
be, however,  borne in mind: DCC can only be produced during 
out-of-equilibrium cooling \footnote{In an adiabatic
process, the domain size is comparable 
to $m_\pi^{-1} \sim T_c^{-1}$
and there is no way of producing a 
coherent multipion state \protect\cite{rw1}.}
and an infinity of distinct
stochastic processes can have in common 
the same equilibrium ensemble. Moreover, one is
here mostly interested in non-universal parameters, 
and not in the behaviour
of the system in the immediate vicinity 
of the phase transition. 
\par
 With the appropriate choice of units, the Lagrangian is 
\begin{equation}
L = {1 \over 2} (\partial \phi)^2 -
{ \lambda \over 4} (\phi^2 - 1)^2 + H \sigma \label{lag}
\end{equation}
The small symmetry breaking term is introduced to take 
into account the effect of small quark masses. Only the 
case $\lambda \gg H$ is relevant for phenomenology. 
Since we are interested in the dynamics of the model, 
a specification of the initial conditions is mandatory. 
Here, we touch another source of uncertainty: One wishes to 
start the evolution in a state with 
unbroken chiral symmetry. But above 
the chiral phase transition point the use of the 
$\sigma$-model is questionable, and at very high temperature, it
presumably  makes no sense !
\par
At this point it is customary to introduce the idea 
of a ``quench'' \cite{rw1}. One assumes that the hot plasma 
is suddenly frozen, and that its subsequent 
dynamics is correctly given by the
(zero temperature) equations of the 
linear $\sigma$-model. In that way the
problem becomes  mathematically well defined, although the difficulty 
has been actually only displaced: What is the 
physical mechanism of the quench? As will be argued later
on, it is likely that a rapid expansion of the system
produces a damping of fluctuations, which is indeed
approximately equivalent to a quench.
\par
Most of the works which have been 
done so far, with the exception of some 
attempts to incorporate quantum corrections 
which we shall discuss in a later
section, deal with the {\em classical} linear 
$\sigma$-model. In particular, the
various ``scenarios'' proposed
refer, in fact, to various approximations 
used to solve the complicated
classical dynamics of the model, 
and to different ways of implementing the
initial conditions. The purpose of the next
section is to give the reader a qualitative
insight into the dynamics of the linear
$\sigma$-model, through selected illustrative examples.

\section{Qualitative trends}
The field equations read
\begin{equation}
\partial^2\phi = - \lambda (\phi^2 - 1)\phi + H n_{\sigma} 
\label{mot}
\end{equation}
where $n_{\sigma}$ is the unit vector in the $\sigma$ 
direction. In this section, quantum effects are neglected, 
so that $\phi$ is to be regarded as a {\it classical} field.
The field $\phi$ evolves in a potential which, apart from
the symmetry breaking term $\propto H$, has the form of a
``Mexican hat''.  We wish to follow the time evolution 
of $\phi$, starting somewhere near the top of the 
``Mexican hat'' and ending at the very bottom. Mathematically, the
classical problem is precisely defined once the initial
conditions $\phi( \bfx ,t)$, 
$\partial_t\phi( \bfx ,t)$ are specified. In its full generality the
problem is fairly complicated. We start by 
adopting an idealization, originally due to Heisenberg \cite{heis}.
This will enable us to follow analytically all the stages 
of the time evolution of $\phi$, taking into account 
the expansion of the system, at the expense of dramatically 
reducing the number of degrees of freedom.

\medskip
\noindent{\bf Time evolution during expansion}\cite{bk2} 
\medskip

Assume that initially, at time $t=0$, the whole energy of 
the collision is localized within an infinitesimally 
thin slab with infinite transverse extent. The symmetry 
of the problem then implies that $\phi$ is a function 
of the {\em proper} time $\tau = \sqrt{t^2 - x^2}$ only. 
The field equations become ordinary differential equations and
can be solved analytically. We shall not enter here into
the algebra, concentrating on the significance of the results.
\par
Of course, viewed in the laboratory, the system expands:
at time $t$ it extends from $x=-t$ to $x=t$. Notice, 
that $\partial^2 \phi \equiv \ddot{\phi} + \dot{\phi}/\tau$, 
where the dot denotes the derivative with respect to 
$\tau$. Hence, the equations of motion involve not 
only the acceleration term but also a friction term. 
This ``friction" reflects the decrease of the energy in a 
covolume, due to expansion, and turns out to be very important .
\par
From eqs. (\ref{mot}) one obtains easily
\begin{equation}
\vec{\pi} \times \dot{\vec{\pi}} = \vec{a}/\tau
\label{vec}
\end{equation}
\noindent
and
\begin{equation}
\vec{\pi}\dot{\sigma} - \sigma \dot{\vec{\pi}} =
\vec{b}/\tau + (H/\tau) \int^{\tau} \vec{\pi} \tau d\tau
\label{axi}
\end{equation}
\noindent
Eq. (\ref{vec}) is a consequence of the conservation 
of the isovector current, while eq. (\ref{axi}) reflects the partial
conservation of the iso-axial-vector current. The isovectors
$\vec{a}$ and $\vec{b}$ are integration constants. The 
lengths of these vectors measure the initial strength of 
the respective current. It is easy to see that the
component of $\vec{\pi}$ along $\vec{a}$ 
vanishes, $\pi_a=0$.
\par
One can show that the second term on the RHS of eq.
(\ref{axi}) is irrelevant as long as $\tau \ll b/\sqrt{H}$
and $a \ll b$. With $H=0$ one can write
\begin{eqnarray}
\pi_b & = & -r \sin{\theta} \\
\pi_c & = & {a \over {\sqrt{a^2 + b^2}}} \; r \cos{\theta} \\
\sigma & = & {b \over {\sqrt{a^2 + b^2}}} \; r \cos{\theta} 
\end{eqnarray}
\noindent
where $\vec{c}=\vec{a} \times \vec{b}$. Thus the motion 
is planar. It remains approximately so even
at later time provided the condition $a \ll b$ is satisfied.
The component $\pi_c$ is then always very small and the
pion field oscillates along the (random) direction defined
by the isovector $\vec{b}$. We concentrate our 
attention on this particularly simple and interesting case.
Assuming that at $\tau=\tau_0$ the distributions of
$\phi$ and $\dot{\phi}$ are Gaussian, 
with variances $\sigma_f$ and $\sigma_g$
respectively, one can calculate the probability 
that $b$ takes a given value:
\begin{equation} 
{{dP(b)} \over {db}} \propto 
{{ e^{- b/b_0}} \over {\sqrt{b}}}
 \; , \; 0 < a < A \ll  b
\label{prob}
\end{equation}
\noindent
where $b_0 = \sigma_f \sigma_g \tau_0$ and $A$ is an 
arbitrary constant. This is an
important result, because it turns out that the initial 
strength $b$ of the axial current controls the time
evolution (see below).
\par
The motion can be described by two variables, the
radial variable $r$ and the angular variable $\theta$. 
Solving the corresponding equations of motion one
can identify several stages in the {\em proper} 
time evolution of this simple dynamical system:
\par
$\bullet$ By assumption, for $\tau < \tau_0$ the model
does not apply. (A plausible value of $\tau_0$ is 1 fm/c.)
\par
$\bullet$ For $\tau_0 < \tau \;
{\mbox{\lower0.6ex\hbox{\vbox{\offinterlineskip 
\hbox{$<$}\vskip1pt\hbox{$\sim$}}}} } \;
b/\sqrt{2\lambda}$ the radial and the angular 
motion are strongly coupled. Both are damped by
friction, which, as we have already explained, is
another facet of the expansion.
\par
$\bullet$ For $ b/\sqrt{2\lambda} \; 
{\mbox{\lower0.6ex\hbox{\vbox{\offinterlineskip 
\hbox{$<$}\vskip1pt\hbox{$\sim$}}}} } \; \tau \;
{\mbox{\lower0.6ex\hbox{\vbox{\offinterlineskip 
\hbox{$<$}\vskip1pt\hbox{$\sim$}}}} } \; b/\sqrt{H}$
the radial motion corresponds to high frequency 
damped oscillations about the equilibrium position 
$r = 1$. For large enough $\tau$ the solution 
takes the particularly transparent form
\begin{equation}
r = 1 + C \cos{(\tau \sqrt{2\lambda} + 
\delta)}/\sqrt{\tau \sqrt{2\lambda}}
\label{rsol}
\end{equation}
\noindent
where $C, \delta$ are constants sensitive to the initial
conditions. Remember, that in this context $\sqrt{2\lambda}$
should be regarded as a large parameter. Setting 
$r \to \langle r \rangle = 1$ in the equation of 
motion for $\theta$, the latter 
becomes that of a damped pendulum. Provided $b$ 
is large enough the motion of the pendulum is circular:
\begin{equation}
\theta = \ln{(\tau/\tau_0)}
\label{tsol}
\end{equation}
\noindent
The regime just described corresponds to the non-linear
$\sigma$-model (we  recover the solution found in
\cite{bk1}).
\par
$\bullet$ Near $\tau \approx b/\sqrt{H}$ the
damping produces a cross-over from the circular to the
oscillatory motion and one finds
\begin{equation}
\pi_b \approx -\theta \approx 
\sqrt{b} \cos{(\tau\sqrt{H} + \delta')}/\sqrt{\tau \sqrt{H}}
\label{ast}
\end{equation}
The RHS of (\ref{ast}) is a solution of the Klein-Gordon 
equation, describing free propagation of pions with mass
$\sqrt{H}$. These are the DCC decay products one hopes to
observe.
\par
It is elementary to find the Fourier 
transform (with respect to $x$) of the RHS
of (\ref{ast}) and to calculate the energy
radiated at large $t$. One finds
a rapidity plateau of height $b$. The plateau is a consequence 
of the boost invariant boundary conditions. The relevant
result is that {\it the energy released in the decay of 
DCC is proportional to the strength of the
initial iso-axial-vector current}. This result
together with (\ref{prob}) suggests that
the probability to release large energy
via DCC decay strongly depends on the 
initial conditions (via the parameter 
$b$) and that it is damped exponentially.
Thus, {\em observable} DCC are likely to be
rare events. If this conclusion is correct, then the
calculation of {\em average}
DCC characteristics is of little
phenomenological interest.

\medskip
\noindent{\bf Amplification of long
 wavelength modes}[11,20-22]
\medskip

With Heisenberg's boundary conditions the problem is
eventually reducible to that of a
dynamical system in $0+1$ dimensions.
But the true problem is $1+3$ dimensional and
its solution requires using a computer.
However,
the most salient conclusions reached via numerical
simulations can be qualitatively understood within an
approximate framework. This is what
we are going to explain now. 
\par
Let us separate the field $\phi(\bfx,t)$ into its spatial
average $\langle\phi(t)\rangle$ and a space 
dependent fluctuating part
$\delta\phi(\bfx,t)$:
\begin{equation}
\phi(\bfx,t)=\langle\phi(t)\rangle+\delta\phi(\bfx,t).
\label{phiaverage}
\end{equation}
We have
\begin{equation}
\langle\phi(t)\rangle=\frac{1}{\Omega}\int {\rm d}^3 x\,\phi(\bfx,t),
\end{equation}
where $\Omega$ is the volume of the 
system and, by definition, the spatial
average of
$\delta\phi(\bfx,t)$ vanishes: 
$\langle\delta\phi\rangle=0$.  
By taking the spatial average of the equation of motion, we get:
\begin{equation}
\partial_t^2\langle\phi(t)\rangle=
- \; \lambda\left( \langle\phi^2(\bfx,t)\phi(\bfx,t)\rangle
-\langle\phi(t)\rangle\right) + Hn_\sigma,
\label{aver}
\end{equation}
where we have assumed that the spatial derivative of 
the field vanishes at the boundary of the reference volume.
By replacing, on the RHS of this equation $\phi(\bfx,t)$ 
by its decomposition (\ref{phiaverage}), one 
obtains an equation which involves spatial 
averages of products of fluctuations, that is, the equation for
$\langle\phi(t)\rangle$ is not a closed one. 
It needs to be complemented
by the equations of motion for the fluctuations.
\par
At this stage, we introduce some {\it approximations} in 
order to treat the fluctuations. 
\par
 (i) Given a product of fluctuation fields, one replaces
all the pair products $\delta\phi_j \delta\phi_k$ 
by $\langle \delta\phi_j 
\delta\phi_k \rangle$, and  one adds  terms 
corresponding to different contraction schemes, much in 
the same way as in writing the well known Wick theorem 
in field theory. Consequently the Gaussian average of the 
initial product of fields and that of the final approximate 
expression are identical. Notice that, with this 
aproximation, the average of a product of an odd 
number of fluctuation fields is zero. 
\par
(ii) We assume that the $4 \times 4$ tensor
$\langle\delta\phi_j\delta\phi_k\rangle$ is diagonal
in the (moving) orthogonal frame whose one axis points
along $\langle \phi \rangle$. 
\par
As a further simplification, we shall also use the 
formal large $N$ limit (of the O(N) $\sigma$-model), neglecting
 terms like 
$\langle \delta\phi_j^2 \rangle$, of the 
order O(1), as compared to $\langle \delta\phi^2 
\rangle$, which is of the order O(N) \footnote{This approximation is 
necessary to fulfill Goldstone theorem  when $H=0$. Otherwise, the
Goldstone bosons acquire a non-vanishing mass, of order $1/N$.}.
\par
The equation for $\langle \phi \rangle$ is immediately 
found from (\ref{aver}). The equations for the
fluctuations are obtained from the exact (classical) 
field equations, by subtracting the
equation of motion for $\langle \phi \rangle$.
Introduce the notation: 
\begin{eqnarray}
\omega^2_\perp(k,t) & = & k^2 + \lambda\left(
\langle \phi(t) \rangle^2+
\langle \delta\phi^2(t) \rangle - 1 \right)\\
\omega^2_\parallel(k,t) & = & k^2 + \lambda\left(
3 \langle \phi(t) \rangle^2+
\langle \delta\phi^2 (t)\rangle -1 \right)
\end{eqnarray}
One easily finds 
\begin{equation}
\partial_t^2\langle\phi(t)\rangle = 
- \; \omega^2_\perp(0,t) \langle\phi(t)\rangle
+Hn_\sigma,
\label{hartree1}
\end{equation}
and
\begin{eqnarray}
\partial_t^2 \delta\phi_\parallel({\bf k},t) & = &
- \; \omega^2_\parallel(k,t) \; \delta \phi_\parallel({\bf k},t)\\
\partial_t^2 \delta\phi_\perp({\bf k},t) & = &
- \; \omega^2_\perp(k,t) \; \delta \phi_\perp({\bf k},t)
\end{eqnarray}
Here $\delta\phi_\parallel$ ($\delta\phi_\perp$) denotes
the fluctuation parallel (perpendicular) to 
$\langle  \phi \rangle$. The approximation 
which we have just done is usually
refered to as the Hartree approximation. One 
encounters a similar structure when discussing quantum 
corrections (see the next section). In the present 
section, the Hartree approximation is introduced 
for purely pedagogical purposes: 
it helps to understand the main features
of the non-linear classical dynamics. 
Solving numerically  eqs. (19)-(21) is
not really  simpler than solving the exact
classical   equations (\ref{mot}).
\par
It is obvious from eqs. (19)-(21) that $\delta\phi$ will be amplified 
during the time evolution whenever the corresponding
$\omega^2$ becomes negative. The following points 
should be clear from the inspection of the formulae
for $\omega$:
\par
$\bullet$ The amplification can only occur if
$k^2$ is small enough. In other words, only long
wavelength modes are amplified.
\par
$\bullet$ The amplification of $\delta\phi_\perp$
stops when $\langle \phi \rangle$ approaches unity. The
amplification of $\delta\phi_\parallel$ is 
not expected to be significant (notice the 
factor of 3 in front of $\langle \phi \rangle^2$ in (18)).
\par
$\bullet$ Large fluctuations prevent the amplification.
Thus one does not expect DCC domains to form when the
energy density (temperature) is too high.
\par
One could derive Hartree equations taking into account expansion, 
as in the earlier example. Assuming that the fields depend
on time via $\tau = (t^2 - \sum_1^D x^2_i)^{1/2}$
one finds a friction force $-(D/\tau)\partial_\tau \phi$. 
Since friction damps fluctuations, we arrive to the conclusion:
\par
$\bullet$ Introducing expansion favors the formation
od DCC domains. The larger is $D$, the stronger is the effect.

\section{Quantitative estimates}
More realistic calculations require
the use of a computer. First
simulations \cite{rw2} have been carried out 
in the static set-up, i.e. neglecting expansion. 
Classical equations of motion have been used, 
assuming (ad hoc) that initially $\phi$ and
$\partial_t\phi$
are Gaussian random variables living on a
discrete grid of points. A dramatic
amplification of long wavelength modes have 
been observed: the larger the wavelength, the 
stronger the amplification. These results 
have been confirmed by other people 
\cite{ggp}.  Notice that 
the amplified soft modes
are not spatially separated from the
hard ones. One should remember this
point when refering to ``DCC domains''. 
Moreover, the amplification
of soft modes is a transient phenomenon: 
The non-linearity 
of the equation of motion leads eventually 
to the equipartition of the energy in a static set-up.
Of course, this does not mean that
DCC signal will not be observed, since, 
in particular,  the expansion may
prevent the system to reach this regime .
\par
The expansion can be introduced 
as explained in the preceding section.
Upgrading the simple model presented there by 
assuming that $\phi$ depends not only on 
the proper time $\tau$ but also on the rapidity variable 
$\eta = {1 \over 2} \ln{{t-x} \over {t+x}}$ it
is found \cite{hw} that DCC is localized in 
a rapidity interval 2 to 3 units long. In
ref. \cite{ahw} the same group has 
assumed invariance under longitudinal boosts,
allowing the system to expand transversally.
The results are very encouraging. With appropriate 
initial conditions the domains of DCC
with 4-5 fm in size have been observed at 
$\tau=5$ fm. But, the emergence of DCC
strongly depends on the choice of the initial conditions 
(see also \cite{bcg}).
\par
A systematic method of sampling initial field
configurations from an equilibrium ensemble
at a given temperature has been devised in 
ref. \cite{ran}. In ref. \cite{ran2} the same
author has studied the dynamical trajectories
generated by the classical equations of motion
(\ref{mot}) starting from initial configurations
generated using the sampling method quoted above.
The trajectories are drawn on the ($\langle \phi \rangle,
\langle \delta\phi^2 \rangle$) plane, where the region
of instability is also exhibited. When the system is 
prepared at the temperature $T$ = 400 MeV the incursion
into the unstable region (i.e. amplification of soft
modes) only occurs for $D > 1$. At $D=3$ instabilities
occur for starting temperatures ranging from 200 MeV to 500 MeV  
(at least). What happens, is that at the initial stage 
of the evolution the fluctuations fall rapidly, while 
$\langle \phi \rangle$, small initially, changes 
little so that the system enters the instability 
region. Then, $\langle \phi \rangle$ 
increases steadily and eventually the 
instability is shut-off. At much higher temperatures the
fluctuations do not have time to decrease enough before
a substantial increase of $\langle \phi \rangle$ and the
instability conditions are never met.
\par
Although DCC is essentially a
quasi-classical phenomenon, several 
quantum effects could play a role.
 First, the classical evolution cannot 
continue forever: when
the energy density becomes low enough, the 
description of the system of particles in 
terms of a classical field becomes meaningless.
There is also another effect 
\cite{krz}, less trivial and usually disregarded
in the DCC context: DCC evolves as an open 
system. The interaction with other nuclear
debris, acting as a ``bath'', is the source 
of decoherence. There are also quantum 
corrections to the dynamics of the condensate.
\par 
Attempts to deal with the last problem 
have been made by several groups. The
authors of ref. \cite{bvh} have considered 
a static set-up, analogous to the one of ref.
\cite{rw2}, quantizing conventionally
on the hypersurfaces $t=$ const. In refs. \cite{ckmp,ckm}
the expansion is taken into account and the 
system is quantized on the hypersurfaces $\tau=$ const. 
Thus the two definitions of the final state wave functions
are not equivalent. Particle production is 
calculated by following the adiabatic vacuum 
of the fluctuation and measuring
particle production with respect to this vacuum. 
In order to proceed both groups 
are led to make approximations and their final 
equations, those used in the actual numerical work, 
have exactly the structure of the Hartree equations written in the 
preceding section. There are obvious modifications:
$\langle ... \rangle$ becomes the average over thermal and
quantum fluctuations and one has to define the product of
fluctuation fields at coincident points, which
requires regularization and renormalization, in the standard
fashion. On the whole it appears that the 
 quantum corrections do not change 
qualitatively the picture, and that the
conclusions of these studies are not in variance with 
expectations drawn from the study of 
the classical equations. Starting with
thermal fluctuations one can hardly 
produce any amplification without
expansion.  And the results are very 
sensitive to initial conditions. Let us
mention that an enhancement at low transverse momentum, 
correlated with the DCC formation has been reported in 
\cite{ckm}.

\section{Lessons from numerical simulations} 
The numerical simulations provide an underpinning to
the qualitative discusion presented in sect. 3.
The main thing that has been learned is that 
in more or less realistic set-ups the initial 
fluctuations can get amplified during the
evolution of the system, giving rise to the 
emergence of a DCC-like phenomenon. Thus, it 
is plausible, but not certain, that the effect will be
occasionally strong enough to be observable. 
The other important lesson is that the results 
depend strongly on the initial conditions.
And, as we have already emphasized, in
the present state of the
art the choice of initial conditions
is the Achilles's tendon of the
theory.
\par
Nobody has been able to estimate
the most relevant parameter: the probability 
of producing an observable DCC signal. We are fully
aware of the fact that the calculations carried
out so far are theorist's games. Their purpose is 
to gauge an idea, not to produce numbers
to be directly compared with experiment.
However, even in a theory that is not fully 
realistic one would be pleased to learn whether 
the effect is expected to occur at the level of one per 
cent or one per billion. A result
analogous to (\ref{prob}),
but valid in a more realistic set-up
would be most welcome.
One can associate a measure with the
thermal configurations of the
$\sigma$-model \cite{ran}. This measure
can, at least in principle,
be used to estimate the probability
we are talking about. Although,
as already mentioned, the use of
the $\sigma$-model is questionable 
above the critical temperature, such
a phase-space measure may
be a good guide. It is not uncommon
in multiparticle production
phenomenology to obtain reasonable
estimates from statistical
arguments, also in instances where,
strictly speaking, thermal
equilibrium arguments do not apply.

\section{Experimental signatures}
There is no doubt that the predicted large 
fluctuations of the neutral
pion fraction are the most striking signature 
of DCC formation. There are, however, at least two
problems in identifying such a signal.
First, one can wonder to 
what extent the simple law
(\ref{law}) is distorted by secondary effects. 
Second, assuming that the distortion is
not very significant, one faces the 
problem of extracting the DCC signal from the
background.
\par
Various corrections to (\ref{law}) have 
been studied in the literature. 
Eq. (\ref{law}) could be modified due to the
coupling of the DCC isospin to the isospin 
of other collision debris. The 
authors of ref. \cite{cbnj} find that
the effect is insignificant, at least for 
$0.1 < f < 0.9$, as long as the isospin
 of DCC is less than 30\% of the pion multiplicity
\footnote{It has been argued, however, that 
isospin non-conserving effects could be
amplified due to coherence,
altering the standard expectations \cite{coh}.}. 
Another source of distortion could come
from final state pion-pion interactions
with charge exchange. But, the mean
free path of soft pions is estimated to
be much larger than the DCC domain size, 
so that the final state interactions 
should not be very important \cite{hs}.
We suspect, that blurring
of (\ref{law}) will mostly come from the pion 
field orientation in isospace being not exactly
constant all over the DCC domain. 
\par
Concerning the problem of DCC signal 
identification, we would like to report 
about an interesting suggestion put forward 
in ref. \cite{hstw}. The authors propose to 
use the modern technique of signal processing, 
the so-called multiresolution wavelet analysis, 
to study the distribution of $f$ on the lego 
plot. Let us briefly explain the idea in one dimension.
\par
 Consider a histogram $H^0$ with bin
size 1 (in appropriate units).
Use it to produce a smoother histogram $H^1$, 
doubling the bin size. Of course,
$H^0 = H^1 + R^1$ and there are
fine structures in $R^1$ over distance 1. 
Repeat the operation to get
$H^0 = H^n + R^1 + ... R^n, \; n = 1,...N$. 
By definition $N$ is such that $H^N$ is structureless. 
The result of these manipulations is a series of 
increasingly coarse-grained histograms:
$H^n$ is living on bins of size $2^n$ and the associated $R^n$ has
fine structures at scale $2^{n-1}$ only. The 
beauty of the story is that $R^n$ can be 
written as a superposition of the so-called wavelet 
functions $W^n_j(x)$, which form a set orthonormal 
with respect to both indices. Furthermore all 
$W^n_j(x)$ can be obtained by rescaling and shifting a ``mother''
function $W(x)$:
\begin{equation}
W^n_j(x) = 2^{-n/2} W(2^{-n} x - j)
\label{wav}
\end{equation}
Various mother functions have been explicitly constructed. 
There exist computer programs performing the 
decomposition \cite{recip}. The extension
to more than one dimension is straightforward. Wavelets are
local in space and scale and are therefore, contrary to
 trigonometric functions, particularly suitable 
to uncover localized structures and find
the associated scales.
\par
The authors of \cite{hstw} have applied this technique
 to analyse rapidity distributions of $f$ 
generated by the classical evolution of the
1+1 dimensional linear $\sigma$-model, in the 
conditions leading to DCC formation. They
compared these ``DCC data'' with ``random noise 
data'' and found very significative
diferences. In particular, the power spectrum
 associated with a given scale
\begin{equation}
P^n = \sum_j \mid (H^0, W^n_j)\mid^2
\label{pow}
\end{equation}
\noindent
has a dramatic scale dependence for ``DCC data'', 
while it is scale independent in the random 
sample. It will be, of course, interesting
to see how the method performs in more realistic cases.
\par
Secondary signatures of DCC 
formation have also been suggested. They 
include specific pion pair correlations
\cite{bd,ggm} and anomalies in electromagnetic 
decays of resonances \cite{hsw}.

\section{Conclusions}
 There exist a few cosmic ray
events, the so-called Centauros 
(see \cite{lfh}), where one observes 
jets consisting of as many as 100 charged 
pions and no neutrals. Are Centauros
an evidence for DCC formation ? We would 
not risk any definite answer. For the
moment, the search for more 
Centauro-like events in cosmic ray interactions
and in accelerator data has been unsuccessful. 
An experiment at the Tevatron
\cite{bt} has been designed to look 
for the phenomenon. Other experiments,
with heavy ions, are being planned and should be encouraged. 
Until the idea does not receive a firm
experimental confirmation it will 
continue having the status of a smart
speculation. However, this speculation 
has led people to think more about
non-equilibrium processes in 
high-energy nuclear collisions and this is
certainly a very positive development. 
Whatever will be the future of this
idea, we have already learned quite 
a lot on the theory side. What is amusing,
is that the discovery of Centauro 
events has been met with widespread
scepticism: how can one seriously 
claim that it is possible to produce a
multipion state with so small a 
fraction of neutrals? Now, as we have a
plausible mechanism for the effect, 
experimenters should be prepared to hear
the opposite blame: the phenomenon is 
so natural, how can it be that you
don't see it?

\end{document}